\documentclass[aps,prb,showpacs,twocolumn,floats,epsfig,pdflatex]{revtex4}
\usepackage{amssymb}
\usepackage{amsbsy}
\usepackage{amsmath}
\usepackage{epsfig}
\usepackage{graphicx}
\begin{document}

\title {Study of unconventional superfluid phases and the phase dynamics in spin-orbit coupled bose system}

\author{Anirban Dutta and Saptarshi Mandal}

\affiliation {Department of Theoretical Physics, Indian Association for
the Cultivation of Science, Jadavpur, Kolkata-700032, India.}

\date{\today}

\begin{abstract}
  We study the phase  distribution and its dynamics in spin-orbit coupled two
component ultracold Bosons for finite size system. Using an inhomogeneous 
meanfield analysis  we demonstrate how  phase 
distribution evolves as  we  tune the spin-orbit coupling $\gamma$ and $t$, 
the spin-independent hopping. For $t>>\gamma$ we find the homogeneous superfluid
phase. As we increase $\gamma$, differences in the phases of the order parameter 
grows leading to twisted superfluid phase. For $t \sim \gamma$  competing orderings
in the phase distribution is seen. At large $\gamma$ limit a Ferro-Magnetic stripe ordering 
appears along the diagonal. We explain that this is due to the frustration bought in by the
spin-orbit interaction. Isolated vortex formation is also shown to appear. We also 
investigate the possible collective modes. In deep superfluid regime 
we derive the equation of motion for the phases following a semi-classical approximation. 
Imaginary  frequencies indicating the damped modes are seen to appear and  the dynamics of 
lowest normal modes are discussed.

\end{abstract}


\maketitle

\section{Introduction}
\label{intro}

The recent advancement in optical lattice experiments to investigate 
the idealized strongly correlated many body system has  initiated a
great interest among the  condensed matter community \cite{morsch}. Starting 
from mimicking simple tight binding Hamiltonian in a periodic lattice, it  can
now  create more complex situations  seen in real materials. Creation of 
artificial abelian or non-abelian gauge fields, density-density 
interaction are some of them to mention \cite{daliberd,linrmp}. Experimental realization of  
Mott-Insulator to Superfluid transition  for ultracold bosons 
\cite{Greiner1,Orzel1} in such system  became a paradigm of itself. 
Recently there has been  experimental realisation to simulate tunable 
spin-orbit coupling in  neutral bosons in optical lattice\cite{sopapers1,ylin} . 
This has been remarkable because it is known that for real material  
spin-orbit coupling is in essential an intrinsic \cite{toprev1} 
properties of the material and  could not be controlled. The
 spin-orbit interaction can  change the physical 
properties of the system dramatically. In optical lattice  the 
spin-orbit coupling is achieved by  Raman laser induced transitions 
between the two internal states of a neutral bosonic atom.
The resulting spin-orbit interaction could be purely  Rashbha \cite{rashbha}
type   or Dresselhaus \cite{dresselhaus} type or suitable combination
of both. 

The result of such spin orbit interaction has been studied
extensively recently \cite{victor1, arun1, cw1, grab-sarma, chunjiwang, ychang}.
In the Mott regime it is shown to support exotic magnetic textures, such as 
vortex crystals and skyrmion lattice \cite{victor1, arun1, cw1}. The 
signature of the Mott-Insulator to Superfluid transition has been shown
to be associated with precursor peaks in momentum distributions 
\cite{saha1, sinha1, issacson1, man-saha-sen}. Various other equilibrium 
and non-equilibrium dynamics has also been analyzed  which could have 
interesting experimental signatures \cite{arun2}.  Boson fractionalisation 
has also been proposed and  formation of twisted superfluid phases has 
been noticed as a result of  spin-orbit interaction \cite{man-saha-sen,panahi}. 
It may be mentioned that for the fermionic case  interesting many body dynamics
has  also been observed  \cite{vi1}.

  The Mott-Insulator to Superfluid transition is well captured by   
Bose-Hubbard model \cite{fisher1,sachdev1,jaksch1}. There are already a large 
number of work done  to investigate the low energy properties 
of such Bose-Hubbard model \cite{sesh1,trivedi1,sengupta1,dupuis1,hrk1}. 
However much of these work was mainly aimed at investigating the systems 
which are thermodynamically large and in weak couple regime. In this work 
we look into the effect of  
spin-orbit interaction of two component bosons  in strong coupling limit
 for different finite size systems. We are motivated to look into  microscopic 
manifestation of the spin-orbit interaction and  various ramifications 
of superfluid order parameter for different system size and  different 
parameter regime. For this we employ  Gutzwiller projected inhomogeneous 
meanfield treatment \cite{hrk1}  which seems to be pertinent for such 
small system size. We work in the strong coupling limit where the Hubbard interaction 
is the highest energy scale of the problem. This limit enables us 
to take the  number of states in the Gutzwiller projected state to be 
necessary minimal. Below we describe our plan of work.

In section I, we begin by giving a detail analysis of the meanfield procedure
and obtain the phase diagram for MI-SF transition. Following this, we  look 
into the phases and magnitude of the SF  order parameter in  superfluid 
regime. We show that the phases and the magnitude of the SF order parameter
respond non-trivially as the parameters are  varied.  We find that when
 $t >> \gamma$, the SF phase is described by a  homogeneous superfluid 
where the magnitude and the phases of the up spins are spatially uniform. 
For intermediate values of $t$ and $\gamma$ we find that the phases and
the amplitudes of both the spins are inhomogeneous and shows interesting 
nontrivial pattern. Depending on the relative strength it could be superposition
of local homogeneous phases and patches where the phases form a  spiralling
pattern. For the limit  $\gamma >> t$, the  phases of the order parameter 
develops a Ferromagnetic order along the diagonal direction followed by periodic 
modulations of magnitude of the SF order parameter. We explain that this is 
due to inherent frustration brought in by the spin-orbit interaction.

In section II, we study the fluctuations  around the meanfield configuration 
 and investigate into the lowest possible excitations. In  section III, 
we study the dynamics of phases inside  the deep SF regime. Assuming that 
the phases of the order parameters are the  low energy degrees of freedom 
in this regime, we deduce the Lagrange-Equation  of motion for  it. 
We find the normal modes.  It appears that due to the constrained
collective motion imaginary frequency appears signifying   damped vibration. 
We also look at the nature of lowest normal modes of the vibrations. 
\begin{figure}
\includegraphics[height=50mm,width=90mm]{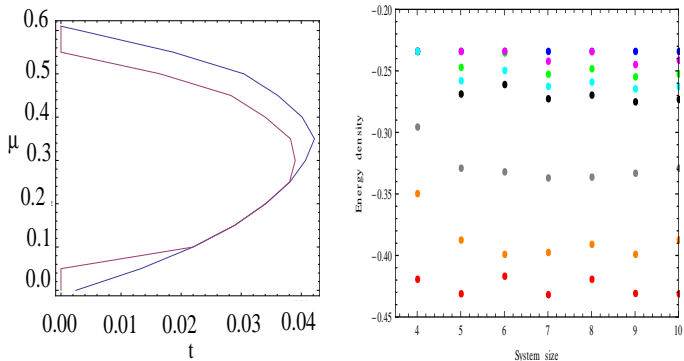}
\caption{In the left panel we have shown the MI-SF transition for $\lambda=0.6, \Omega=0.01$ in $\mu-t$ 
plane. The blue line denotes the transition for $\gamma=0$ and the red line denotes $\gamma=0.04$.
In the right panel we have plotted the energy density per plaquette with various system sizes for
different parameter values. The various color represents various set of ($\gamma, t$) which are
explained in the text.} \label{fig0}
\end{figure}
\section{Meanfield study}
As already mentioned in this work we study a spin-orbit coupled two component bosons
in square lattice. The Hamiltonian for such a system can be written 
as $H=H_0 + H_1$, where $H_0$ 
and $H_1$ are given by \cite{ylin,man-saha-sen},
\begin{eqnarray}
&&H_0= \sum_{ia} -\mu n_{ia} + U n_{ia}(n_{ia}-1) + 
\lambda U \sum_i n_{i1} n_{i2} \nonumber\\
&& - \sum_{\langle ij \rangle_a} t_{a} b^{\dagger}_{ia} b_{ja}, 
~~~H_1= i \gamma \sum_i \Psi^{\dagger}_i \hat{z} . 
(\vec{\sigma} \times \vec{d}_{ij}) \Psi_j \nonumber \\
&& + \sum_i \left( \delta \Psi^{\dagger}_i \sigma_y \Psi_i - 
\Omega  \Psi^{\dagger}_i \sigma_z \Psi_i\right) \label{ham1}
\end{eqnarray}
Here $\Psi_i= \left(b_{i1}, b_{i2} \right)$.
In the above Hamiltonian $\mu$ represents the chemical potential, $\Omega$ is
the Zeeman shift between the two species, $U$ is the intraspecies 
interaction and $\lambda$ is the on site interspecies interaction. For 
the meanfield analysis we take the Gutzwiller variational wave function 
$|\Psi \rangle= \prod_i |\psi_i \rangle$, where $|\psi_i\rangle$ is the 
wave function at a given site '$i$'. $|\psi_i\rangle$ is given by 
$|\psi_i \rangle = \sum_{m,n} f_{m,n} |mn \rangle$. As we  work in a 
strong coupling limit where $U$ is much larger that $t$ and $\gamma$, 
it is sufficient to take states upto 2 particle at
a given site. The meanfield order parameter is defined as, 
$\Delta_{ia}= \langle \psi_i | b_{i a} | \psi_i \rangle$. The expression 
for $\Delta_{ia}$ in terms of $f_{mn,i}$ are given below,
\begin{eqnarray}
&&\Delta_{i1}= f_{10,i} f^*_{00,i} + f_{11,i} f^{*}_{01,i} + \sqrt{2} f_{20,i} f^*_{10,i} \nonumber \\
&&\Delta_{i2}= f_{01,i} f^*_{00,i} + f_{11,i} f^{*}_{10,i} + \sqrt{2} f_{02,i} f^*_{01,i} \label{ord}
\end{eqnarray}

The first part of the Hamiltonian in Eq.(\ref{ham1}) contains the on site interactions 
and we call  it $H_{at}$ which given by,
\begin{eqnarray}
&\langle H_{at} \rangle_i= - \mu_{1} \Big( |f_{10,i}|^2 + 
|f_{11,i}|^2 + 2 |f_{20,i}|^2 \Big) + \lambda U |f_{11,i}|^2-&
\nonumber \\
&\mu_{2} \Big( |f_{01,i}|^2 + |f_{11,i}|^2 + 2 |f_{02,i}|^2 \Big)
+ 2U \Big( |f_{20,i}|^2 + |f_{02,i}|^2 \Big)~~~& \label{hat}
\end{eqnarray}
A generic term in $H_1$  can be written as $b^{\dagger}_{i, \alpha} b_{j,\beta}$. 
The meanfield decomposition of it is given by,
\begin{eqnarray}
b^{\dagger}_{i, \alpha} b_{j,\beta}=\Delta^{*}_{i\alpha} b_{j\beta} + 
\Delta_{j\alpha} b^{\dagger}_{i\beta}- \Delta^{*}_{i\alpha} \Delta_{j\beta} \label{del11}
\end{eqnarray}
After we substitute Eq.(\ref{hat}), Eq.(\ref{del11}) in Eq.(\ref{ham1})  and use  Eq.(\ref{ord})  
we  can write the meanfield decomposed Hamiltonian as,
\begin{eqnarray}
&&H=\sum_i \chi^{\dagger}_i   F_i(\mu, \lambda, \Delta_{j,\alpha}, t, \gamma) \chi_i \label{mham}
\end{eqnarray}
where $\chi_i= \left( f_{00,i}, f_{10,i}, f_{01,i}, f_{11,i}, f_{20,i}, f_{02,i}\right)$. 
The problem then reduces to diagonalizing  the matrix $F_i$ at every site self consistently. The 
Hamiltonian in Eq.(\ref{mham} is  still a coupled problem. We notice that 
in the presence of spin-orbit coupling $\Delta_{i}$ can not be taken uniform at each 
site for then the  spin-orbit interaction contribute nothing to the total energy. To find 
the meanfield solution, we start from a given random initial distribution of $\Delta_{i}$ 
at each site $i$ diagonalize the $F_i(\mu, \lambda, \Delta_{j,\alpha}, t, \gamma)$ at each site.
We then calculate the new set of $\tilde{\Delta}_i$ corresponding to the minimum eigenvalue
of $F_i$. The resulting $\tilde{\Delta}_i$'s are fed back into Eq.(\ref{mham}) until $\Delta_{i}$
 becomes equals to $\tilde{\Delta}_i$ at each site $i$. We do this procedure for approximately 
$10^4$  random configurations and take the configurations of $\tilde{\Delta}_{i}$ which 
corresponds to the global minima. In the Fig.(\ref{fig0}) left panel, we show the phase diagram 
for the MI-SF transitions. In the right panel of Fig.(\ref{fig0}), we have plotted the 
energy density per site with the system size for various set of parameter . 
We find that finite size minimization brings significant variations in the energy 
density with the system size. The various color represents various set of parameters 
($\gamma, t$). Red represents (0.1,0.02), blue represents (0.02,0.04), green represents 
(0.03,0.04), black is for (0.04,0.04), gray is for (0.06,0.04), orange denotes (0.08,0.04) 
magenta denotes (0.025,0.04) and cyan is for (0.035,0.04). This color scheme is maintained for 
all the figures that will be used later. In the following we discuss the textures of the 
order parameter $\Delta_{ia}$ for different values of $t$ and $\gamma$.
\vspace{-0.5cm}
\subsection{Numerical results}
 First we discuss the regime when  $t >> \gamma$ followed by the 
regime where $t \sim \gamma$. Lastly we discuss the regime where 
$\gamma >> t$.
\begin{figure}
\includegraphics[height=50mm,width=80mm]{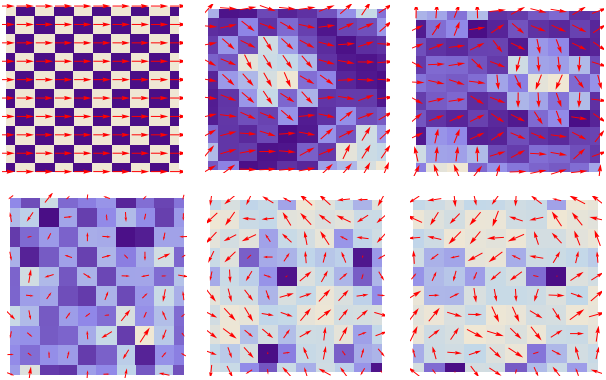}
\caption{Magnitude and phase of the order parameter is plotted at
each site. The arrows represent the phases and the color represent the
magnitude of the order parameter $\Delta_{i}$. The upper panel denotes
phase and magnitude for $\Delta_{1}$ and the lower panels are for $\Delta_2$.
The left panels denotes the result for $\gamma=0.02, t=0.04$. The middle
panels are for $\gamma=0.025, t=0.04$ and the right panels are for $\gamma=0.03, t=0.04$.  
 } \label{fig1}
\end{figure}
\begin{figure}
\includegraphics[height=50mm,width=80mm]{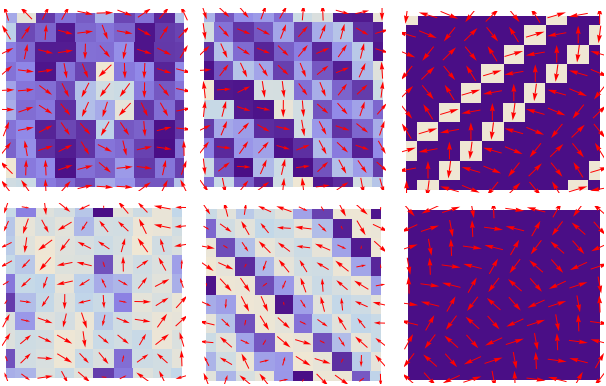}
\caption{ The distribution of phases and the order parameter as explained in Fig.(\ref{fig1}). Here the
left panels represent the result for $\gamma=0.035, t=0.04$, the middle panels
represent $\gamma=0.04, t=0.04$ and the right panels represent $\gamma=0.06, t=0.04$.} \label{fig2}
\end{figure}
\vspace{-0.5 cm}
\subsubsection{Meanfield results when $t$ is large compared to $\gamma$.}
In Fig.(\ref{fig1}) we present the resulting  distributions of phases and the 
magnitude of the order parameter $\Delta$. The arrows represents the phases 
and the background color represents the relative magnitudes of the order 
parameters. The dark color represents greater magnitude. The upper panel 
is for $\Delta_1$ and the lower panel is for $\Delta_2$. In the extreme 
left panel the result is shown for $t=0.04, \gamma=0.02$. We find that 
the distribution of phases $\Delta_1$ are ordered and spatially
uniform while that of $\Delta_{2}$ is disordered. The magnitudes of $\Delta_1$
shown  form a two sublattice structure, however there are degenerate solutions
with spatially uniform magnitude. It is clear that the two sublattice structure
is the result of spin-orbit interaction.  Also we have $\langle \Delta_{1} \rangle 
>> \langle \Delta_{2} \rangle$. The above textures is understood easily as for 
the presence of $\Omega$, the system is favoring the condensation of species 1 
which resembles the homogeneous superfluid. The middle panel of Fig.(\ref{fig1}), 
represents the result for $t=0.04, \gamma=0.03$.  We observe that the phases 
are no longer uniform leading to twisted superfluid phase ~\cite{panahi}. 
We observe the reduction of the ordered pattern of $\Delta_1$ and onset of 
diagonal  ordering. The  magnitude of $\Delta_1$ are also random. The $\Delta_2$ also 
shows signature of  diagonal ordering. The competition of ordering along the two 
diagonal shows the signature of large vortices as seen in the moddlelower panal in 
Fig.(\ref{fig1}).
\vspace{-0.5 cm}
\subsubsection{$t$ and $\gamma$ is comparable}

The phase textures for this regime could be described as follows. We find a 
competition between local ferro magnetic alignment for nearest $\Delta_i$'s 
and the ferromagnetic(FM) ordering along the  diagonal neighbors. The FM 
ordering for the neighbors results from  direct hopping. Where as the 
ferromagnetic ordering along the diagonal is due to the spin-orbit 
coupling as explained in next section. In Fig.(\ref{fig2}), left panel 
represents the phase distribution for $\gamma=0.035, t=0.04$, middle panel 
is for $\gamma=0.04, t=0.04$ and the right panel is for $\gamma=0.06$. 
We notice that the minimum energy configuration presented here is not unique. 
There are many degenerate  configurations with identical energy. However
the quantum fluctuations would pick the global minima. For examples, 
in  Fig.(\ref{fig2}), we find the onset of density modulations and no vertex
formations. There are degenerate meanfield solutions with completely
random density distribution with isolated vertex formations.
\begin{figure}
\includegraphics[height=50mm,width=80mm]{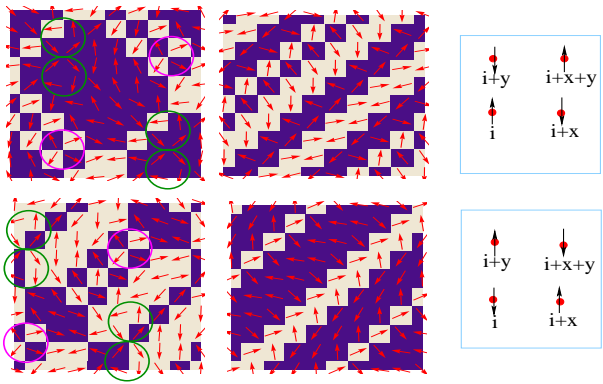}
\caption{The distribution of phases and the order parameter as explained in 
Fig.(\ref{fig1}) and (\ref{fig2}). The left and the middle panal is drawn 
for $\gamma=0.1, t=0.02$. These represent the degenerate meanfield configuration. 
We see that the left panal contains vertex and antivertex. The green circle
contains the vertex configurations and the  pink circle contains antivertex. 
In the right panel we have shown the spin-orbit coupled hopping processes 
for up-spin starting from site $i$ in anti-clockwise direction.} \label{fig3}
\end{figure}
\vspace{-0.5 cm}
\subsubsection{$t$ is small and $\gamma$ is large}
In this regime we notice that the phases forms a ferromagnetic alignment
along the diagonal. The magnitude of the order parameter are also
seen to be modulated. In Fig.({\ref{fig3}}), we present the distribution for
the phases and the magnitude of order parameter for $\gamma=0.1, t=0.02$. We
see that ferromagnetic ordering of phases along the diagonal is common. While
in the left panel FM ordering happens for both the diagonal, for the middle 
panel it happens for only (1,1) direction. In the left panel isolated vertex 
\cite{victor1}  and anti vertex is seen to appear. To understand the  
phase distribution in this regime it may be 
useful to consider an elementary square plaquette and consider 
the meanfield  Hamiltonian for it. Let us consider 
the hopping of an up spin under spin-orbit coupling via the  sites 
$i$, $i+x$, $i+x+y$ and $i+y$ in anti-clockwise direction as shown in the right upper 
panel in fig \ref{fig3}. The meanfield decomposition put the following 
constraints on the phases,
\begin{eqnarray}
\theta_{i,1}- \theta_{i+x,2}= \pm \pi,~~ 
\theta_{i+x,2}-\theta_{i+x+y,1}=\frac{\pi}{2}, \nonumber \\
\theta_{i+x+y,1}-\theta_{i+y,2}=0,~~~
\theta_{i+y,2}-\theta_{i,1}=-\frac{\pi}{2}
\end{eqnarray}
The above set of equations do not have simultaneous solutions for all the parameters.
One may eleminate $\theta_{i+x,2}$ (and $\theta_{i+y,2}$ ) from the   
1st and 2nd (and 3rd and 4th) to solve for $\theta_{i,1}$ and 
$\theta_{i+x+y,1}$ to obtain that they are equal,  the numerical outcome
seems to conform  this. It then poses an ill-defined equation for 
 $\theta_{i+x,2}$ (and $\theta_{i+y,2}$ ) which is fixed to 
minimize the  plaquette energy. The ratio of average plaquette energy 
obtained from  numerics to that obtained by minimizing a single 
plaquette is 0.94 which is satisfactory. In recapitulation we have
shown within meanfield how the twisted superfluid phase appears as we 
gradually tune the parameter $t$ and $\gamma$ for a tight binding 
Hamiltonian. We have shown the onset of density modulations and stripe 
pattern ~\cite{ho-wang} for the phases as the $\gamma$ is increased gradually.

\section{Fluctuation around the meanfield}
Here we look into the fluctuations around the meanfield solutions obtained
in the previous section.  To take into the role of  fluctuation we expand 
Gutzwiller coefficients ~\cite{constantin} $f_{mn,i}$ around its saddle point 
and expand it by $f_{mn,i} =\bar{f}_{mn,i} + \delta f_{mn,i}$ where $\bar{f}_{mn,i}$ 
represents the equilibrium values. After we  substitute it in  Eq.(\ref{mham}) we  
retain the terms which are quadratic in $\delta f_{mn,i}$
(and its complex conjugate). The resulting Hamiltonian then could be written as,
\begin{eqnarray}
H= \Psi^{\dagger} H_{\delta} \Psi
\end{eqnarray}
where $\Psi=(\psi_1, \psi_2,...\psi_r,....\psi_N)$ and $\psi_i=
\big( \psi_{ui}, \psi_{di} )$.  Here $\psi_{ui}=(\delta f_{00,i} ~\delta f_{10,i} 
~\delta f_{01,i}~  \delta f_{11,i}~  \delta f_{20,i}~  \delta f_{02,i})$ and 
$\psi_{di}= \psi^{*}_{ui}$. It is clear that $H_{\delta}$ represents a 
$12 N \times 12 N$ Hermitian matrix whose eigenvalues and eigenvectors represents 
the collective modes. It may be noted that the substitution, $\delta f_{mn,i}= 
\sum_k u_{mn,k} e^{i k r} + v_{mn,k} e^{-i k r}$, does not simplify the problem 
as the $\bar{f}_{mn,i}$'s are not translational  invariant.  We denote the eigenvalue closer 
to  absolute zero by $E_0$. The $E_0$  is a measure of possible Goldstone modes 
of the system and is shown in \label{fig78}. We find that for $t>> \gamma$, the system  always 
find zero energy modes. For $t \sim \gamma$, where the phases are disordered we 
also find similar behavior. However for $\gamma >> t$, we find that  $E_0$ is
$\sim 10^4$ times larger than the other parameter regime. However the $E_0$ scales to lower values
monotonically as we  increase the system size. The gradual decrease of $E_0$ with 
system size $N$ indicates that it is approaching to possible zero energy modes.
The reason that $E_0$  for $\gamma >> t$ is larger than other cases by few thousand
order is the following. For $t>>\gamma$ the uniform phase distribution always find
Goldstone modes and there is no frustration in the system also. For $t \sim \gamma$,
the spins are disordered and random. Thus it is  easily possible to re-distribute
the phases to have zero energy eigenmodes which is nearly degenerate with the 
original solutions. However for $\gamma >> t$, the distribution of phases and the
magnitudes are governed by the  frustration bought in by spin-orbit
coupling. The degenerate solutions in this case as seen from Fig.(\ref{fig3}) are
not easily connected. Thus the collective excitations  costs finite energy
than the other cases. However as we increase the system size, we expect that the 
degenerate  solutions are easily obtained from one other leading to zero energy
mode.  We also observe that the eigenvalues of the collective modes 
form three distinct bands. This is clear  from Eq.(\ref{hat}). The fluctuation of $f_{2,0}$ 
or $f_{0,2}$ yields the bands around  $U$. While the fluctuation of $f_{11}$ yields the 
bands around $\lambda/2$. The fluctuation of $f_{1,0}, f_{0,1}$ and $f_{0,0}$ constitutes 
the lower bands.  We denote these three bands by $E_2, E_1$ and $E_0$ respectively. 
In the right panel of Fig.(\ref{fig78}), we have plotted the band-width with the system 
sizes for different parameter values. In the left panel of Fig.(\ref{fig87}) we have plotted the 
bandwidth of $E_1$ and the right panel is for $E_0$. It appears that for a given $t$, 
the bandwidth is inversely proportional to $\gamma$. Also more the value of $\gamma$, 
the bandwidth oscillates more with the system sizes. We notice that the bands $E_2$ and 
$E_1$ are symmetric but $E_0$ is not because of the presence of $\Omega$.

\begin{figure}
\includegraphics[height=30mm,width=80mm]{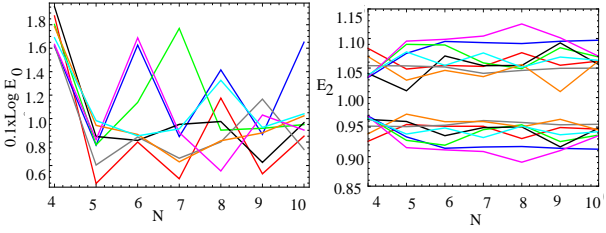}
\caption{In the left a measure of zero energy eigenmodes $E_0$ 
due to the collective motions has been shown. In the right panel 
the bandwidth around $E_2$ has been plotted. For both the figure
horizontal axis represnts the length a $N\times N$ lattice. The 
various colors represent various set of parameters as given in the text.} \label{fig78}
\end{figure}
\begin{figure}
\includegraphics[height=30mm,width=80mm]{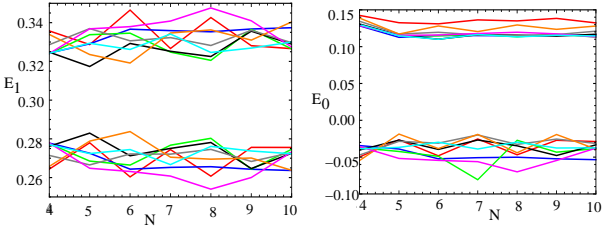}
\caption{In the left panel the bandwidth around $E_1$ has been shown. The 
right panel is for $E_0$. In both the figure the horizontal  axis represents
the length of a $N\times N$ lattice. } \label{fig87}
\end{figure}

\section{Dynamics of the phases} 

Now we turn out attention to the deep inside the superfluid regime where
one may neglect the fluctuations of the magnitude of the order parameter
and consider the phases as the only relevant degree of freedom. 
Following a semi-classical approximation, we deduce the  Lagrangian
and the equation of motion for the phases and determine the normal modes of the 
vibrations.  The meanfield decomposition of Eq.(\ref{ham1}) could be
written as,
\begin{eqnarray}
&&H= \sum_i  \mu_{\alpha,i}\langle n_{\alpha,i} \rangle +\frac{U}{2} 
\langle n_{\alpha,i} \rangle^2  + \lambda U  \langle n_{\alpha, i} \rangle  
\langle n_{\beta,i} \rangle \nonumber \\ &&~~~~ -  \sum_{\langle ij \rangle} \left( \lambda_{ij, 
\alpha\beta} \Delta^*_{\alpha,i}  \Delta_{\beta,j} +h.c \right)~~~~~ \label{hon}
\end{eqnarray}

\begin{figure}
\includegraphics[height=50mm,width=90mm]{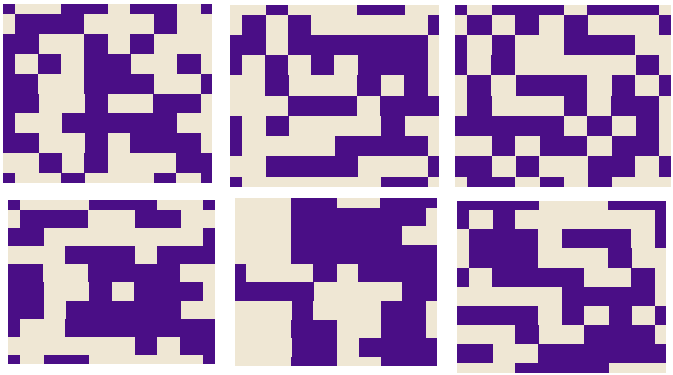}
\caption{We have shown the nature of vibration for the lowest normal modes. In each
panel the upper panales denotes $\theta_{1}$ and the lower panels are for $\theta_2$.
The right panel denotes the the case $\gamma <t$, ($\gamma=0.02, t=0.04$). The middle
panal reprensents  $\gamma \sim t$, ($\gamma=0.04, t=0.04$). The right panel represent 
$\gamma >t$, ($\gamma=0.1, t=0.02$). The white region denotes the motion in clockwise
direction and the blue region denotes motion in the anti-clockwise direction.} \label{fig9}
\end{figure}

In the above $\lambda_{ij,\alpha\beta}$ denotes a general hopping parameter.
The main  disadvantage  of   Eq. (\ref{hon}) is that  all the 
variables commute with each other and bear no signature of the  original 
bosonic commutation relations. To derive Lagrangian of the phases of 
the order parameter $\Delta$, we follow the procedure in \cite{legget,kivelson}. 
Translating the original  bosonic commutators to the commutation relations 
of the meanfield variables, we find that,
\begin{eqnarray}
[n_1, b_1]=-b_1  \rightarrow [\langle n_1 \rangle, \Delta_1 ]= - \Delta_1 \label{n1b1} 
\end{eqnarray}
Writing $\Delta_1 = e^{i \theta_1} |\Delta_1|$ and keeping $|\Delta_1|$ constant we obtain,
\begin{eqnarray}
[\langle n_1 \rangle, e^{i \theta_1} ]= - e^{i \theta_1}
\end{eqnarray}
Expanding $e^{i \theta_1}$ and keeping only the lowest order term we obtain 
for $\theta_1 \rightarrow 0$, the following commutation relations,
\begin{eqnarray}
[\langle n_1 \rangle, \theta_1 ]= i,~~[\langle n_1 \rangle^2, \theta_1 ]=
2 i \langle n_1 \rangle \label{cmt}
\end{eqnarray}
The above procedure yields the following coupled equations to be solved for 
the $\frac{\partial \theta_{\alpha}}{\partial t}$ and $\frac{\langle n_{\alpha i} 
\rangle }{\partial t} $
\begin{eqnarray}
&&\frac{\partial \theta_{1i}}{\partial t} = - (\mu + \Omega + \frac{U}{2}) +  
U \langle n_{1i} \rangle + \lambda U \langle n_{2i} \rangle \nonumber \\
&&\frac{\partial \theta_{2i}}{\partial t}= - (\mu - \Omega + \frac{U}{2}) + 
 U \langle n_{2i} \rangle +\lambda U \langle n_{1i} \rangle
\end{eqnarray}
Solving for  $\langle n_{1i} \rangle$ and $\langle n_{2i} \rangle$ from the above 
two equations  and substituting in the Hamiltonian, Eq. \ref{hon}, we obtain the following equations,
\begin{eqnarray}
&&H_{sh}= B_0 \left( (\frac{\partial \theta_{1i}}{\partial t} )^2 +
 (\frac{\partial \theta_{2i}}{\partial t} )^2 \right) + B_1 \frac{\partial 
\theta_{1i}}{\partial t} +  B_2 \frac{\partial \theta_{2i}}{\partial t} \nonumber \\
&&+ B_3  \frac{\partial \theta_{1i}}{\partial t} \frac{\partial \theta_{2i}}
{\partial t} +F(\theta_{i1},\theta_{i2})+ B_4 
\end{eqnarray}
Where $F(\theta_{i1},\theta_{i2})$ is given in the appendix. Expressions for $A_i$'s are
also  given in the appendix. To derive the E-L equations of motion, we  introduce 
the relative and total phase by the relation, $\theta_{1i}= \theta_{ci} + 
\theta_{ir},~~~\theta_{2i}= \theta_{ci} - \theta_{ir} $
After inserting the above change of variables we can rewrite Eq ~\ref{lagh} as follows,
\begin{eqnarray}
&&H_{n}=\sum_i T_1 (\dot{\theta}_{ic} + \alpha_c)^2 + T_2 (\dot{\theta}_{ir} + \alpha_r)^2 \nonumber \\
&& ~~~~+ F(\theta_{ir}, \theta_{ic}) + \sum \alpha_{icr}
\end{eqnarray}
Here $T_{1/2}= 2 B_0 \pm B_3$
Using the above equations, we write the resulting Lagrangian and the equation
of motion below,
\begin{eqnarray}
&&\mathcal{L}= \sum_iT_1(\dot{\theta}_{ic} + \alpha_c)^2 +  T_2 (\dot{\theta}_{ir} + 
\alpha_r)^2 - F(\theta_{ir}, \theta_{ic})\nonumber \\
&&\ddot{\theta}_{ic}=- \frac{\partial F(\theta_{ic}, \theta_{ir})}{T_1\partial \theta_{ic}},
~~\ddot{\theta}_{ir}= -\frac{\partial F(\theta_{ic}, \theta_{ir})}{T_2\partial \theta_{ir}}
\label{eleq}
\end{eqnarray}
In the last equation we have deliberately omitted the inconsequential constant term 
$\sum \alpha_{icr}$. After simplifying the r.h.s of Eq.\ref{eleq} and subsequently expanding 
upto linear term we can  rewrite it is, $\ddot{\Theta}= M\Theta$. Where for a system of 
$N \times N$ lattice $\Theta$ is a column matrix with $2N^2$ element such that 
$\Theta_{i}= \theta_{ic}$ and $\Theta_{N^2+i}=\theta_{ir}$ where $i$ runs from 
1 to $N^2$. $M$ is a $2N^2 \times 2N^2$ matrix. The eigenvalues of the 
matrix $M$ yields the normal modes. We find that the due to the presence of $\gamma$, 
the normal modes develop negative  eigenvalues signifying damped modes. 
In fig \ref{fig9} we have plotted schematically  the lowest normal modes for 
three different regime. In all the plot the blue region denotes displacements  of
phases in forward direction (anti-clockwise rotation) and the white regions denotes 
displacements in the backward directions (clockwise rotation). The right panel 
denotes the case for $\gamma>>t$, the middle panel denotes  $\gamma \sim t$ and 
the right panel is for $t>> \gamma$. In each of these panel the upper one 
denotes the displacement for species 1 and the lower panel describe the 
displacements for species 2. Looking at the upper panel we find that for the 
$\gamma >>t$, there is tendency of phases to move synchronously along the diagonal 
which is expected. However for the middle panel and the left panel there is a 
preferences in horizontal ordering and patches of areas  vibrating in  breathing modes.
For the species 2, the left panel, we find similar behavior though region executing breathing modes
are larger. 

\section{Discussion}
To summarize we have explored the  different phases that might occur for a
spin-orbit coupled bosons in the optical lattice. We have  extensively
studied the distribution of phases and the magnitude of order parameter for 
varying finite size system  using an inhomogeneous meanfield analysis. We
have shown that for a given $t$, as we increase the spin-orbit interaction
$\gamma$,  we observe the destruction of normal homogeneous superfluid  phase and 
onset of twisted superfluid phases. At large $\gamma$ limit  an interesting
ordering along the diagonal appears. We have also investigated the  fluctuation
around the meanfield and shows the existence of Goldstone modes. The scaling of
minimum energy excitations with system size has also  been shown. Finally, using
semiclassical approximation we derived the equation of motion for the phases and
derive the normal modes of vibrations.   We think that some of the results 
may have interesting experimental signatures in the light of recent  experiments.

\vspace{-0.5cm}
\section{appendix}
\vspace{-0.75cm}
\begin{eqnarray}
b_0&=&  \frac{\lambda^2_0 u}{2},~~b_3= u \lambda \lambda^2_0 \left(1- \frac{1}{\lambda^2} \right)\nonumber \\
 \nonumber \\
b_1&=&  \lambda_0 (\frac{a_1}{\lambda}-a_2) + \lambda_0 u (a_2 -\frac{a_1}{\lambda}) + u \lambda \lambda_0 (a_1 -\frac{a_2}{\lambda}) ~~~~~~~\nonumber \\
b_2&=&  \lambda_0 (\frac{a_2}{\lambda}-a_1) + \lambda_0 u (a_1 -\frac{a_2}{\lambda}) + u \lambda \lambda_0 (a_2 -\frac{a_2}{\lambda}) ~~~~~~~\nonumber \\
b_4&=& -a_1 a_1 - a_2 a_2 + \frac{u}{2} (a^2_1 + a^2_2) + \lambda u a_1 a_2 \nonumber \\
a_1&=& \mu+ \omega+ \frac{u}{2},~~~~~a_2= \mu - \omega+ \frac{u}{2},~~~\lambda_0= \frac{\lambda}{u(\lambda^2-1)}\nonumber \\
a_1&=& \lambda_0 \left(a_2 - \frac{a_1}{\lambda} \right),~a_2= \lambda_0 \left(a_1 - \frac{a_2}{\lambda} \right),\end{eqnarray}
\begin{eqnarray}
&&\alpha_c= \frac{b_1+ b_2}{2(2b_0+b_3)}, ~~\alpha_r= \frac{b_1- b_2}{2(2b_0-b_3)}, ~\alpha_{icr}=-\alpha^2_c-\alpha^2_r
\nonumber \\
&&f(\theta_{i1},\theta_{i2})\nonumber \\
&&=-2 \gamma_t \sum_{\langle ij \rangle} \left( \cos(\theta_{1i}- \theta_{1j}) + \eta \beta^2 \cos(\theta_{2i}- \theta_{2j}) \right) |\delta^2_1| \nonumber \\
&&- 2 \gamma_s \beta \sum_{\langle ij \rangle_x} \left( \cos(\theta_{2i}- \theta_{1jx}) - \cos(\theta_{1i}- \theta_{2jx}) \right) |\delta^2_1| \nonumber \\
&&+ 2 \gamma_s \beta \sum_{\langle ij \rangle_y} \left( \sin(\theta_{1i}- \theta_{2jy}) + \sin(\theta_{2i}- \theta_{1jy})\right) |\delta^2_1|
\label{lagh}
\end{eqnarray}
\end{document}